World Scientific
www.worldscientific.com

# Unconditionally Secure, Wireless-Wired Ground–Satellite–Ground Communication Networks Utilizing Classical & Quantum Noise

Lucas Truax *,‡, Sandip Roy † and Laszlo B. Kish †

*Stephenson Stellar Corporation
Shreveport, Louisiana, USA

†Department of Electrical and Computer Engineering
Texas A&M University TAMUS 3128,
College Station, Texas, USA

‡ltruax@stephensonstellar.org

In this paper, we introduce the Kirchhoff-Law-Johnson-Noise (KLJN) as an approach to securing satellite communications. KLJN has the potential to revolutionize satellite communication security through its combination of simplicity, cost-effectiveness, and resilience with unconditional security. Unlike quantum key distribution (QKD), which requires complex, fragile, and expensive infrastructure like photon detectors and dedicated optical links, KLJN operates using standard electronic components and wires, significantly reducing implementation costs and logistical hurdles. KLJN's security, grounded in the fundamental laws of classical physics, is impervious to environmental and radiation-induced noise, making it highly reliable in the harsh conditions of satellite communications. This robustness, coupled with its ability to integrate seamlessly with existing infrastructure, positions KLJN as a revolutionary alternative to quantum solutions for ensuring secure, resilient satellite communications.

The authors explore the value of achieving unconditionally secure communications in strategic ground-to-satellite networks which address vulnerabilities posed by advanced computational threats, including quantum computing. Our team has examined two leading approaches to unconditional security — the KLJN scheme and QKD — and analyzed the potential use of each for space systems. While QKD leverages quantum mechanics for security, it faces challenges related to cost, complexity, and environmental sensitivity. In contrast, the KLJN scheme utilizes classical physics principles to provide a simpler, more cost-effective, and resilient alternative, particularly for ground-based systems. Further, this paper highlights the potential for complementary roles of these technologies in hybrid networks: KLJN can facilitate secure key exchanges without QKD for ground communications connected via wire, and QKD could facilitate secure inter-island and direct-to-satellite communications. Through a detailed analysis of implementation challenges, key exchange speeds, and network architectures, this study concludes that KLJN offers significant advantages in simplicity, cost-efficiency, and robustness, making it a practical choice for many secure communication applications.

## 1. Introduction

In this paper, we examine the necessity for unconditionally secure communications in networks of strategic importance and methods by which unconditional security may be achieved using state-of-the-art methods that are robust against exponential computational power, such as quantum computing (QC). Section 2 discusses the need and hardware methods to achieve unconditional security in remote communications. Section 3 explores several network types as examples. Finally, Sec. 4 discusses speed and related issues. Section 5 presents our conclusion.

## 2. The Need for Unconditional Security in Ground–Ground and Ground–Satellite–Ground Communications

The United States and its allies rely heavily on space assets to achieve critical defense operations with significant future defense capabilities hinged on effective and secure space communications. Critical national defense missions include hypersonic missile tracking, space-based battle management, and "assured, resilient, low-latency military data, and connectivity worldwide to the full range of warfighter platforms [20]." As space-based missions continue to evolve in strategic significance, the security of space communications must be trusted.

Further demonstrating the necessity for unconditional (that is, information-theoretical) security [1–3] in ground–ground (GG) and ground–satellite–ground (GSG) networks is the growing sophistication of cyber threats and the potential vulnerabilities introduced by future advancements in QC [4] and/or noise-based logic (NBL) [5] Rapid advancements in computing power and recent development of post-quantum cryptography (PQC) algorithms [6] offer only conditional security, which will prove only temporarily effective. Unconditional security offers the same features as post-quantum encryption while ensuring that the information exchanged between parties remains confidential and tamper-proof, even against adversaries with unlimited computational power.

QC and NBL present considerable vulnerabilities to secure communications, mainly due to their capabilities to undermine traditional encryption methods. Threats against secure communications include the following.

(a) *Breaking classical encryption algorithms* Scalable, full implementation of QC algorithms on future quantum hardware would efficiently solve mathematical problems that underpin many classical encryption schemes. Notably, Shor's algorithm would allow quantum computers to factor large integers quickly, rendering widely used public key cryptography methods such as RSA and Diffie–Hellman insecure. This capability means that sensitive data encrypted with these algorithms could be decrypted by adversaries using quantum technology, compromising the confidentiality of communications across various sectors, including finance and government.

(b) *Data breaches* QC could enable adversaries to decrypt sensitive information such as personal data, financial transactions, and classified communications, leading to severe privacy violations and economic impacts.

(c) *Cybersecurity landscape shift* The introduction of QC can shift the focus from exploiting human or technical vulnerabilities to directly attacking cryptographic systems themselves. This change requires a reevaluation of current cybersecurity strategies and defenses, long considered a trusted protection for data.

(d) *Transition challenges to quantum-resistant cryptography* As quantum threats loom, there is a strong effort to transition to PQC [6–9], which employs classical algorithms utilizing (*currently*) hard-to-solve mathematical problems that are believed to be resistant to quantum attacks. In other words, PQC offers conditional security only. However, these new algorithms are not yet universally standardized or proven secure against all potential quantum attacks [7–10]. The implementation of PQC involves significant infrastructural changes and investments, which can be challenging for many organizations. Recent advancements in cryptography have revealed vulnerabilities in several post-quantum algorithms, including those recommended by NIST. Some examples are shown as follows.

(e) *CRYSTALS-Kyber PCQ algorithm* The CRYSTALS-Kyber encryption mechanism, which was recommended by NIST for PQC, was successfully broken using a combination of artificial intelligence (AI) and side-channel attacks. Researchers from KTH Royal Institute of Technology utilized deep learning techniques (such as timing and power consumption) to analyze side-channel data in order to extract encrypted keys from the implementation. This attack highlights the potential threats posed by both quantum and classical computing advancements, particularly AI [9].

(f) *Supersingular Isogeny Key Encapsulation (SIKE) PCQ algorithm* This post-quantum algorithm was cracked within an hour using a standard PC [10]. The algorithm was a finalist in the NIST PQC competition, which aimed to identify encryption methods resistant to quantum attacks. The attack utilized mathematical techniques rather than exploiting software vulnerabilities, demonstrating that even legacy hardware could effectively compromise what was considered a secure encryption method. The researchers reported that they could break specific challenges related to SIKE in just a few minutes, indicating the algorithm's significant vulnerabilities despite its intended quantum resistance.

(g) *Implications of these breaks* (see also [11]) The successful attacks on CRYSTALS-Kyber and SIKE raise critical questions about the robustness of post-quantum cryptographic standards. These advancements strongly suggest that, in a high-security network, achieving unconditionally secure communications — necessitating hardware solutions — is likely indispensable. In contrast, the U.S. has increasingly favored software-based security frameworks over robust hardware solutions that could offer superior protection against sophisticated threats. As the National Security Agency (NSA) shifts toward these emerging methodologies, other nations may enhance their own security capabilities, potentially surpassing the U.S. in developing comprehensive hardware solutions for secure communications.

Countries such as China and Russia are making substantial investments in advanced encryption technologies and hardware-based security systems, which could bring significant advantages to national security and cybersecurity. This trend suggests that while the U.S. adapts to a reactive cybersecurity model, it risks lagging in proactive measures that ensure unconditional security. The reliance on software solutions may expose critical infrastructure to vulnerabilities that hardware approaches could effectively mitigate. As adversaries develop increasingly sophisticated attack vectors, the lack of robust, unconditional security measures could render U.S. networks vulnerable to breaches that compromise sensitive information and national security.

### 2.1. *Hardware tools to reach unconditional security in communications*

Currently, there are two main hardware-based approaches that offer an unconditional level of security: the Kirchhoff-Law-Johnson-Noise (KLJN) scheme and quantum key distribution (QKD).

(h) *KLJN scheme* The KLJN scheme [12, 13], which uses classical physics principles, operates over standard electrical wires. It exploits enhanced thermal noise in resistors to generate secure random bits that can be used as encryption keys. This method is based on the laws of classical statistical physics and does not require quantum effects, making it potentially more robust and easier to implement in existing infrastructure. The physical foundation of the unconditional security is the Second Law of Thermodynamics.

The core KLJN system is shown in Fig. 1. The key exchange process involves two identical pairs of resistors, $R_L$ and $R_H$, (with enhanced Johnson noise by external Gaussian voltage generators, one for Alice and one for Bob). At the beginning of each secure bit exchange cycle, the switches connect one randomly chosen resistor to the line at both ends, resulting in four possible states: LL, LH, HL, and HH. The mixed states (LH and HL) exhibit identical effective noise voltages in the line, making it impossible for an eavesdropper, Eve, to distinguish between them. However, Alice and Bob can deduce each other's resistor states based on their knowledge of the channel's noise levels and their own resistor values. They publicly agree how to map the LH and Hl situations to the values (0 or 1) of key bits.

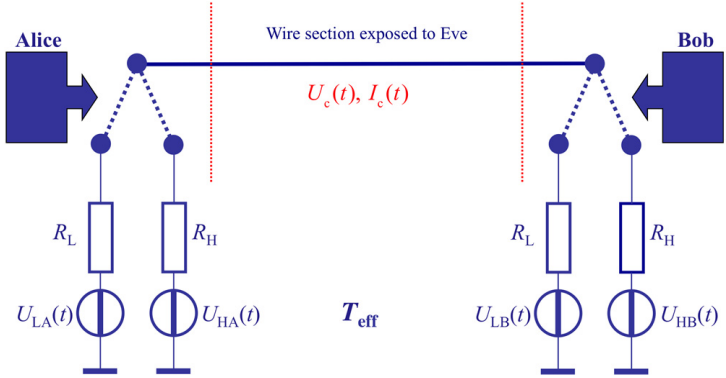

Fig. 1. The core of KLJN secure key exchanger.

The practical system is more sophisticated due to extra circuitry to eliminate information leak with non-ideal elements and active attacks [12, 13].

(i) *QKD* QKD [14], on the other hand, leverages the principles of quantum mechanics — such as the quantum no-cloning theorem — to achieve unconditional security. It can be implemented using two main physical channels:

- Optical fibers: QKD over optical fibers has seen significant advancements, with current experiments reaching distances exceeding 400 km and generating key rates of a few Mbit/s over metropolitan distances.
- Free-space/satellite: To overcome the distance limitations of fiber-based QKD, researchers have been exploring satellite-based QKD. This approach allows for global-scale quantum networks by using satellites to distribute entangled photons or to serve as trusted nodes for key exchange.

Both KLJN and QKD offer theoretical unconditional security, meaning their security is not based on computational complexity but on fundamental laws of physics. While QKD has seen more widespread research and development, the KLJN scheme presents an alternative that is more readily integrated with existing communication systems. In addition, it can be integrated on a chip, which is not feasible with QKD.

### 2.2. *Challenges of QKD and KLJN*

#### 2.2.1. *QKD challenges*

QKD has several significant implementation challenges. They include complexity of implementation, transmission losses, environmental factors increasing error rates and noise, single-photon sources, and a high cost of infrastructure and equipment.

Complexity of implementation — the most significant challenge to quantum-based solutions is the complexity and costliness of implementations. Challenges in this area include the following:

- Quantum approaches require dedicated optical fibers or free-space optical links for photon transmission, which are expensive and impractical over long distances or in challenging environments.
- Integrating quantum cryptographic systems with existing classical communication infrastructure requires significant modifications, posing compatibility challenges.
- Producing reliable single-photon emitters is technologically challenging. Practical systems often use weak laser pulses, which can emit multiple photons, making them susceptible to photon number splitting attacks. Decoy state methods have been developed to mitigate this issue, but they add complexity to the system.
- Implementing quantum-based cryptography requires specialized hardware including specialized transmitters and receivers, and often requires leasing dedicated fiber connections or managing free-space optical links. This hardware dependency complicates integration into existing network infrastructures and limits flexibility for future upgrades or security patches.

*High cost of equipment and infrastructure requirements* — Implementing QKD requires specialized equipment including photon sources, detectors, and stable transmission media. This makes QKD expensive compared with classical cryptography solutions, limiting its appeal for widespread use.

*Transmission losses* — Quantum signals are highly sensitive, and even small transmission losses over long distances can disrupt the quantum states. Fiber optics exponentially attenuates signals over distance, and free-space transmission is limited by weather and atmospheric conditions, making long-distance QKD challenging.

*Error rates and noise* — Quantum states are easily disturbed by environmental noise, which leads to errors in the transmitted data. For example, QKD systems — particularly those using polarization encosing — are sensitive to fluctuations in polarization caused by environmental factors such as vibration of the fibers. This sensitivity arises primarily due to random birefringence in the optical fibers, which can be induced by mechanical stress, temperature changes, and other disturbances. High error rates make it difficult to distinguish legitimate keys from noise, affecting system reliability.

*Single-photon generation* — Ideal QKD requires single-photon sources to prevent eavesdropping attacks, but creating and detecting single photons with high efficiency is technically challenging. Imperfections in photon sources can lead to vulnerabilities. To avoid these problems, the *decoy-state method* has been developed, which is designed to enhance security by mitigating vulnerabilities associated with multiphoton emissions from weak coherent sources. However, several open problems persist within this framework: distinguishability of states by the eavesdropper, noisy channels, hardware complexity and accuracy, correlations between photon packages, etc. On the other hand, the decoy-state method allows high key bit exchange range, even tens of kilobit/second speeds over hundreds of kilometer distances, but a hundred times less via satellites.

*Peer-to-peer (P2P) key exchange scheme* — The fundamental step of key exchange occurs between two parties that need to have a QKD device and an independent optical cable between them. The same situation exists for KLJN. The implications are both hardware and time complexity (speed), (see Sec. 4).

*Sensitivity to radiation* — The single photon detectors are sensitive to radiation, particularly to gamma radiation. On the other hand, neutron radiation — such as that around nuclear reactors — damages the photodiodes. The impacts of neutron radiation on photodiodes include the following:

- Reduction in photocurrent: Studies have shown that neutron radiation induces displacement defects in the semiconductor lattice of photodiodes, resulting in a decrease in photocurrent. For instance, a study on commercial optoelectronic devices, including photodiodes, found that neutron irradiation led to a significant reduction in photocurrent due to the introduction of defects that hinder carrier mobility and recombination processes.
- Increased dark current: Neutron irradiation also tends to increase the dark current in avalanche photodiodes. This increase is attributed to the generation of new carrier centers that contribute to noise, which can obscure the signal needed for QKD applications. The dark current and photocurrent can become indistinguishable under high neutron fluence, indicating a loss of detection capability.
- Effects on gain characteristics: The gain characteristics of Si- and InP-based APDs are negatively impacted by neutron radiation. The gain decreases significantly after exposure, further complicating the ability to detect weak signals necessary for effective QKD. The soft breakdown voltage of these devices also can shift, leading to reduced operational reliability.
- Photocurrent generation mechanism: Photocurrent in photodiodes is generated when photons create electron–hole pairs, especially when absorbed near the depletion region. However, neutron radiation alters this process by introducing defects that affect carrier generation and lifetime, ultimately reducing the efficiency of photocurrent generation.

While neutron radiation does not eliminate photocurrent generation in photodiodes used for QKD, it significantly degrades its performance by reducing photocurrent and increasing dark current. This degradation poses challenges for maintaining reliable operation in radiation-rich environments.

Additional note: There are strong limits of radiation hardening a photodiode because the photons must be able to enter the device. At the same point, gamma radiation and neutrons also can enter.

On the other hand, when no wire or fiber connection is possible (such as with satellites), QKD is the sole solution to share unconditionally secure keys. It is important to note that cosmic radiation has a profound impact on QKD by increasing error rates, which challenges the reliability of quantum communication systems. Ongoing research and development of mitigation strategies are essential for advancing the robustness of satellite QKD technologies.

### 2.2.2. *KLJN challenges*

KLJN has significantly fewer and smaller challenges to implementation than QKD. Challenges with the KLJN approach include the following:

*Transmission losses* — The operational conditions of KLJN must remain in the no-wave (quasi-stationary) limit of frequency range versus distance. In QKD, that implies an exponential slowdown with distance, while in KLJN that implies only a polynomial slowdown [15].

*Speed* — The same condition (and the fact that the bit exchange uses statistics) limits the speed of KLJN to the range of a few hundred bits/second over a kilometer distance. If a KLJN chip is developed and used, then cables with multiple parallel wires can be used (there are phone cables with up to 500 parallel, screened wires) which are driven by chips and multiply the speed accordingly. If the KLJN system is used within a laptop computer that is made unconditionally secure by securing all the communications between the various units in the computer, the hardware speed is a limiting factor, which can provide a high key bit rate.

*P2P key exchange scheme* — Similar to QKD, the fundamental step of key exchange occurs between two parties — they both need to have a KLJN device and an independent wire (possible pair) between them. The implications are both hardware and time complexity (speed) (see Sec. 4). The KLJN challenge is significantly reduced due to the broad range of materials able to be used in KLJN and the ability to integrate a chip.

*Commercialization* — While QKD has already been commercialized, KLJN has not yet reached that stage. There are still additional R&D steps required to identify additional (unknown) KLJN challenges prior to production deployments.

## 2.3. *Benefits of the KLJN approach versus quantum-based approaches*

We turn now to the benefits of the KLJN approach over current approaches to quantum cryptography. The major challenges for quantum-based cryptography have been outlined in Sec. 2.2.1. The most significant of areas of improvement for these issues (when leveraging the KLJN approach) are outlined as follows.

### 2.3.1. *Simplicity and cost-effectiveness*

- Compared with quantum-based approaches, the KLJN protocol uses simple electronic components including resistors, voltage sources, chips, and common wire types. The simplicity of the technology drastically reduces the complexity of the implementation. The combination of known components and low-complexity technology results in KLJN being relatively inexpensive to implement compared to QKD systems.

- *The cost of equipment and infrastructure requirements* is drastically reduced as soon as the KLJN chip is integrated. Residual costs are only the wiring infrastructure.
- KLJN does not require advanced infrastructure (i.e., switched polarizer setups or advanced photon detectors.
- The KLJN method operates at low frequencies and its chip version requires minimal power, making it highly energy-efficient compared with quantum cryptographic methods or other sophisticated physical-layer security techniques.

### 2.3.2. *Enhanced resilience*

Unlike quantum-based systems, KLJN systems are not sensitive to environmental conditions such as temperature fluctuations, vibrations, or other stress-induced polarization fluctuations. Quantum systems are highly sensitive to noise and other perturbations, including their optical fibers. Security solutions must be increasingly resilient against a variety of threats, and the classical physical robustness of KLJN provides significant benefits over quantum approaches.

### 2.3.3. *Uncompromised security*

KLJN provides benefits in each of these areas while also maintaining the strengths of a quantum-based approach, including the following:

- No Assumptions Regarding Computational Power — Unlike cryptographic systems that assume bounded computational power for adversaries, KLJN does not rely on such assumptions. Its security is derived from physical laws that apply universally.
- Unconditional Security — The security of the KLJN scheme is based on a fundamental law of physics, namely the Second Law of Thermodynamics: the impossibility of constructing a perpetual motion machine of the second kind. In other words, it is as impossible to crack the security of the ideal KLJN scheme by a passive (listening) attack as it is to construct a perpetual motion machine.
- Intrusion Detection — Any attempt by an eavesdropper to launch an active attack will change the balance of the thermal noise voltages and currents in the KLJN channel, making the intrusion detectable by legitimate parties.

### 2.3.4. *Extreme robustness against radiation*

Until the chips remain functional against radiation damage, the KLJN system remains operational. Radiation-hardened electronic components are readily

*L. Truax, S. Roy & L. B. Kish*

available. Noises caused by the radiation in semiconductor elements are negligible compared to the macroscopic signal levels in the KLJN line and circuits.

## 3. Examples for Secure GG and GSG Networks

In Figs. 2–4, we introduce generic network situations applicable to KLJN. We group the stations into "islands." Within an island, the stations can communicate with each other via wired (wire or optical fiber) or wireless channels (radio waves). Between the islands, only satellite-based communications are possible.

Figure 2 shows an example of two islands without unconditionally secure communications — that is, without unconditionally secure key exchange between the stations (indicated by black color). The black triangles represent stations without KLJN or QKD. Similarly, the black stars represent satellite communicators

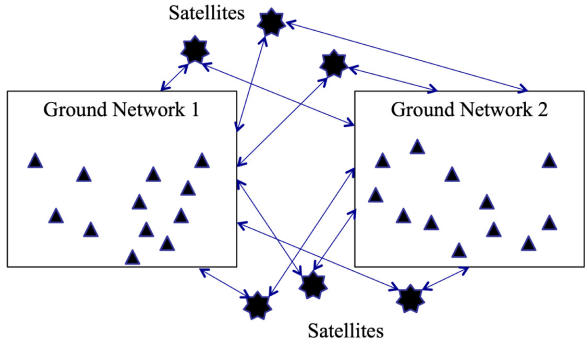

Fig. 2. Conditionally secure GSG network "islands" (such as on different continents) where the islands are connected by satellites with conditional security.

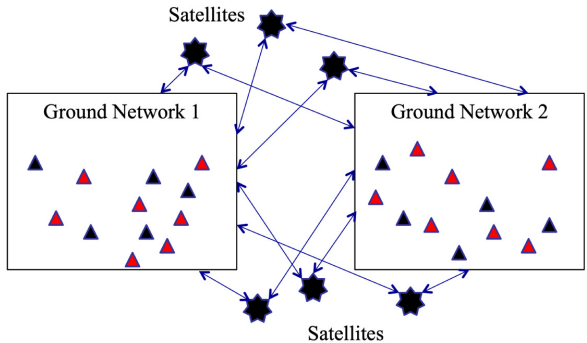

Fig. 3. Islands with GSG networks where some of the stations (red triangles) form unconditionally secure networks. Those stations can communicate unconditionally securely both via ground and satellite connections. However, the satellite connection between the islands is only conditionally secure. Thus, unconditionally secure stations in the two separate islands cannot communicate with each other with unconditional security due to the lack of a properly shared key.

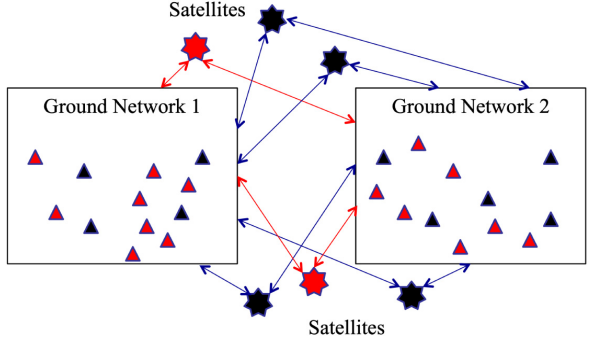

Fig. 4. Islands with GSG networks where some of the stations form unconditionally secure networks and some of the satellite connections between the islands also are unconditionally secured by QKD. In this case, the unconditionally secured members in the two islands can communicate with unconditional security.

*Note.* It is possible to rank the trust of the security of these stations with a straightforward formula [16]. The trust function [16] is a monotonic, saturating function of the number of KLJN and wireless connection of a station.

without QKD. The black arrows represent conditionally secure or unsecured GSG communication channels.

In Fig. 3, some of the ground stations (red triangles) have unconditionally secure key exchangers and can perform unconditionally secure communications with the connected similar units within the islands. However, the satellites are still without QKD, so there are no unconditionally secure communications between the islands. KLJN can provide unconditional security in this situation if keys are shared on the ground via KLJN-secured channels.

Finally, Fig. 4 shows an example where some of the inter-island communication satellites (red stars) are equipped by QKD. That allows unconditionally secure communications between the two islands. The corresponding ground station must also be equipped with QKD, but the key shared via the satellite also can be distributed among the wire-connected stations via KLJN, which makes the inter-island communications unconditionally secure between these connected stations.

## 4. Ground Network Wiring Geometry and Related Key Exchange Speed Considerations

Due to the P2P nature of unconditionally secure key exchangers, the selection of the network geometry is not a trivial question, as it impacts hardware and time complexity (costs and speed) [17]. In this section, we focus on KLJN-based ground networks due to their benefits.

Hardware complexity in this context signifies the number of KLJN units and their connecting lines needed for the network. Time complexity in this context means the average time requirement of the key exchange between the network elements. These two complexities can be traded between each other.

For example, the fastest network is the one where each station is connected with all the other stations [17]. That means, for a network of $N$ stations, each station must have $N-1$ parallel KLJN units; each one is wired to the KLJN unit of a different station. For the whole network, that means that the number of $M$ simultaneously operating KLJN devices needed is

$$M = N(N-1), \tag{1}$$

while the number $W$ of wire connections in the network is

$$W = \frac{N(N-1)}{2}. \tag{2}$$

This is a very high hardware complexity, however the time complexity is excellent: just a single bit exchange period (BEP) is enough to arm the entire network with an unconditionally secure key bit.

In the KLJN scheme, the BEP and the key exchange speed can be estimated as follows [15, 18].

- The KLJN noise bandwidth can be about 10% of the lowest frequency that causes standing waves in the wire.
- The BEP is required to allow a statistical evaluation of the mean-square noise and needs about 100 independent samples.
- Taking into account the conditions, the best-case scenario for exchanging a single bit with the standard KLJN requires a BEP of $\tau_1$ of

$$\tau_1 \approx 2000 \frac{L}{c}, \tag{3}$$

where $L$ is the length of the wire between the two KLJN units and $c$ is the velocity of waves in the cable (about $2 \times 10^8$ m/s).

Thus, a key exchange for $K$ long keys in the network requires a time $\tau_K$ of:

$$\tau_K \approx 2000 \frac{KL}{c}. \tag{4}$$

According to Eq. (4), in such a network with KLJN units located at an average distance of 1 km, the loading of the entire network with all the $N(N-1)/2$ different 256 bit-long unconditionally secure keys requires about 2.5 s, independently from the number of stations $N$.

Such a high hardware complexity can be reduced by trading in increased time complexity [17]. For example, a star network with a central switching exchange (see Fig. 5) where the switching exchange simultaneously connects pairs of stations. Then, the above time complexity increases to $(N-1)\tau_K$. The hardware complexity decreases to $N$ as each station needs only one KLJN unit and one wire connection toward the exchange [17].

Finally, we mention a generic line network [19] where each station has typically two KLJN units (one to "left" and another one to "right"), see Fig. 6. These are the least efficient ones regarding time complexity; however, they do not need an exchange. Integrating chip technology increases the speed of the solution. Phone cables can contain up to 500 screened twisted wire pairs. Such cables driven by chips with a sufficiently large number of integrated KLJN units provide speed enhancement by the same factor.

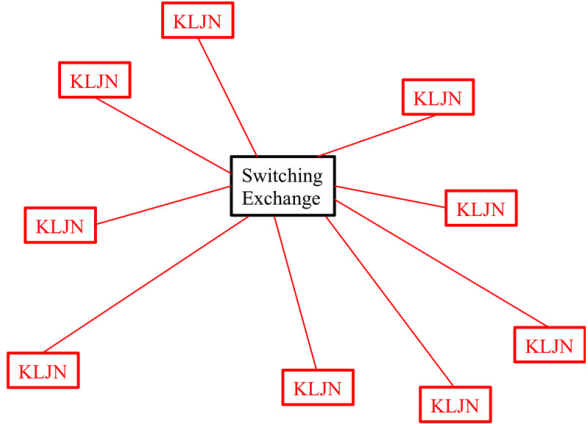

Fig. 5.   Star network: the most economical structure regarding hardware and time complexity.

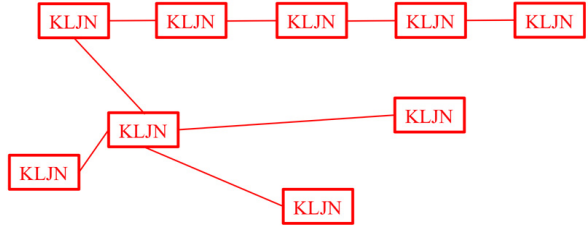

Fig. 6.   Example for an arbitrary combination of a line network with a station with multiple connected units.

## 5. Conclusion

In this paper, we have explored various aspects of creating unconditionally secure data communication networks, where the critical component of unconditional security is the secure key exchange. The KLJN scheme offers substantial advantages over quantum-based approaches for achieving unconditionally secure communications, particularly in ground-based systems. Its reliance on classical physics principles ensures simplicity, cost-effectiveness and resilience, avoiding the significant complexity and environmental sensitivity of QKD. KLJN operates with standard electrical components and infrastructure, reducing implementation barriers and energy requirements while maintaining uncompromised security through the Second Law of Thermodynamics. Its inherent robustness against environmental factors such as temperature fluctuations, noise, and radiation underscores its suitability for critical applications in secure communication networks. If satellite communication is used within an island where the stations are connected within a KLJN key exchange ground network, then high-speed satellite communications can be carried out without quantum-based encryption.

Further, KLJN can be integrated into satellite communication networks to create hybrid systems that leverage the strengths of both KLJN and QKD. By utilizing KLJN for secure key exchange within ground networks and complementing it with QKD for inter-island and satellite-based communications, the framework can ensure global-scale unconditional security. To fully realize these benefits, further research and prototype development are necessary to address practical implementation challenges, optimize performance, and validate the system's effectiveness in real-world environments. The development of KLJN-based hardware and testing in simulated and operational conditions will be essential for transitioning this promising technology from theory to practice, ensuring robust and secure communication infrastructures for future strategic needs.

## ORCID

Lucas Truax  https://orcid.org/0009-0004-3234-0015
Sandip Roy  https://orcid.org/0000-0002-5558-9698
Laszlo B. Kish  https://orcid.org/0000-0002-8917-954X